\documentclass[12pt,preprint]{aastex}
\usepackage{graphicx}
\usepackage{emulateapj5}
\usepackage{apjfonts}
\usepackage{epsfig}
\usepackage{natbib}
\def\gtrsim{\mathrel{\hbox{\rlap{\hbox{\lower4pt\hbox{$\sim$}}}\hbox{\raise2pt\hbox{$>$}}}}}

\newcommand{\kms}{km~s\ensuremath{^{-1}}}

\newcommand{\lamr}{\ensuremath{\lambda_{\rm R}}}
\newcommand{\lamout}{\ensuremath{\lambda_{\rm out}}}
\newcommand{\lamre}{\ensuremath{\lambda_{\rm R_e}}}

\newcommand{\msun}{\ensuremath{M_{\odot}}}

\def\lax{{$\mathrel{\hbox{\rlap{\hbox{\lower4pt\hbox{$\sim$}}}\hbox{$<$}}}$}}
\def\gax{{$\mathrel{\hbox{\rlap{\hbox{\lower4pt\hbox{$\sim$}}}\hbox{$>$}}}$}}

\begin{document}

\title{Probing the kinematic morphology-density relation of early-type galaxies with MaNGA}
\author{J. E. Greene\altaffilmark{1}, A. Leauthaud\altaffilmark{2}, E. Emsellem\altaffilmark{3}, D. Goddard\altaffilmark{4}, J. Ge\altaffilmark{5}, B. H. Andrews\altaffilmark{6}, J. Brinkman\altaffilmark{7}, J.~R.~Brownstein\altaffilmark{8}, J. Greco\altaffilmark{1}, D.~Law\altaffilmark{9},Y.-T. Lin\altaffilmark{10}, K. L. Masters\altaffilmark{4}, M. Merrifield\altaffilmark{11}, S. More\altaffilmark{12}, N. Okabe\altaffilmark{13,14,15}, D.~P.~Schneider\altaffilmark{16,17}, D. Thomas\altaffilmark{4}, D.~A.~Wake\altaffilmark{18,19}, R. Yan\altaffilmark{20}, N. Drory\altaffilmark{20}}
\altaffiltext{1}{Department of Astrophysics, Princeton University, Princeton, NJ 08540, USA}
\altaffiltext{2}{Department of Astronomy and Astrophysics, University of California, Santa Cruz, 1156 High Street, Santa Cruz, CA 95064, USA}
\altaffiltext{3}{European Southern Observatory, Karl-Schwarzschild-Str. 2, D-85741 Garching, Germany}
\altaffiltext{4}{Institute of Cosmology and Gravitation, University of Portsmouth, Dennis Sciama Building, Burnaby Road, Portsmouth PO1 3FX, UK; South East Physics Network}
\altaffiltext{5}{National Astronomical Observatories, Chinese Academy of Sciences, 20A Datun Road, Chaoyang District, Beijing 100012, China}
\altaffiltext{6}{PITT PACC, Department of Physics and Astronomy, University of Pittsburgh, Pittsburgh, PA 15260, USA}
\altaffiltext{7}{Apache Point Observatory, P.O. Box 59, Sunspot, NM 88349}
\altaffiltext{8}{Department of Physics and Astronomy, University of Utah, 115 S. 1400 E., Salt Lake City, UT 84112, USA}
\altaffiltext{9}{Space Telescope Science Institute, 3700 San Martin Drive, Baltimore, MD 21218, USA}
\altaffiltext{10}{Institute of Astronomy and Astrophysics, Academia Sinica, Taipei 10617, Taiwan}
\altaffiltext{11}{School of Physics and Astronomy, The University of Nottingham, University Park, Nottingham, NG7 2RD, UK}
\altaffiltext{12}{Kavli Institute for the Physics and Mathematics of the Universe (WPI), Tokyo Institutes for Advanced Study, The University of Tokyo, 5-1-5 Kashiwanoha, Kashiwa-shi, Chiba, 277-8583, Japan}
\altaffiltext{13}{Department of Physical Science, Hiroshima University, 1-3-1 Kagamiyama, Higashi-Hiroshima, Hiroshima 739-8526, Japan}
\altaffiltext{14}{Hiroshima Astrophysical Science Center, Hiroshima University, Higashi-Hiroshima, Kagamiyama 1-3-1, 739-8526, Japan}
\altaffiltext{15}{Core Research for Energetic Universe, Hiroshima University, 1-3-1, Kagamiyama, Higashi-Hiroshima, Hiroshima 739-8526, Japan}
\altaffiltext{16}{Department of Astronomy and Astrophysics, The Pennsylvania State University,University Park, PA 16802}
\altaffiltext{17}{Institute for Gravitation and the Cosmos, The Pennsylvania State University, University Park, PA 16802}
\altaffiltext{18}{School of Physical Sciences, The Open University, Milton Keynes,  MK7 6AA, UK}
\altaffiltext{19}{Department of Physics, University of North Carolina, Asheville, NC 28804, USA}
\altaffiltext{20}{Department of Physics and Astronomy, University of Kentucky, 505 Rose Street, Lexington, KY 40506-0057, USA}
\altaffiltext{21}{McDonald Observatory, The University of Texas at Austin, 1 University Station, Austin, TX 78712, USA}


\maketitle

\begin{abstract}
The ``kinematic'' morphology-density relation for early-type galaxies posits that those galaxies with low angular momentum are preferentially found in the highest-density regions of the universe. We use a large sample of galaxy groups with halo masses $10^{12.5} < M_{\rm halo} < 10^{14.5} h^{-1}\, M_{\odot}$ observed with the Mapping Nearby Galaxies at APO (MaNGA) survey to examine whether there is a correlation between local environment and rotational support that is independent of stellar mass. We find no compelling evidence for a relationship between the angular momentum content of early-type galaxies and either local overdensity or radial position within the group at fixed stellar mass.
\end{abstract}

\section{Introduction}

The angular momentum content of galaxies can serve as a probe of their assembly histories. Although early-type galaxies are dynamically hot systems, many of them show some rotation \citep[e.g.,][]{daviesetal1983,franxillingworth1990}. The evolution in angular momentum of galaxies is influenced in complex ways by mergers (both major or minor), gas accretion, and internal processes such as star formation that either turn gas into stars or expel gas \citep[e.g.,][]{penoyreetal2017}. Mergers, for instance, might increase or decrease the angular momentum depending on the configuration \citep[e.g.,][]{naabetal2014}. Tracking the evolution in angular momentum content with galaxy properties is a tracer of the factors that dominate galaxy evolution. 

It is standard to trace the galaxy angular momentum using the ratio of velocity to dispersion support. In the context of integral-field spectroscopy (IFS), a luminosity-weighted two-dimensional measurement ($\lambda_R$) is used as a proxy for the angular momentum content.  We adopt the definition of $\lambda_R$ from \citet[][see also \citealt{binney2005}]{emsellemetal2007}, with $R$ the flux-weighted radial coordinate, $V$ the radial velocity, and $\sigma$ the stellar velocity dispersion: $\lambda_R = \langle R | V | \rangle/\langle R \sqrt{V^2 + \sigma^2} \rangle$. Slowly rotating galaxies are those that fall below the expectations for a mildly anisotropic oblate rotator \citep[see][for details]{emsellemetal2007,cappellarietal2006}. Stellar mass is the primary determinant of whether a galaxy is a slow rotator \citep[][]{emsellemetal2007,emsellemetal2011,vealeetal2017a,olivaetal2017}. Secondary correlations with environment may reveal the physical processes that determine the distribution of angular momentum in galaxies.

Early studies examined the relationship between angular momentum as traced by $\lambda_R$ and environment by measuring the fraction of galaxies with low \lamr\ as a function of local overdensity. \citet{cappellarietal2011} showed that a tiny fraction ($<5\%$) of early-type galaxies in low-density environments are slow rotators, and tied this low fraction to a kinematic version of the morphology-density relation \citep{dressler1980}. The fraction of early-type slowly rotating galaxies indeed rises dramatically in the densest environments \citep[][although see also \citealt{fogartyetal2014}]{houghtonetal2013,deugenioetal2015,scottetal2014}. As examples, we present measurements from \citet{cappellarietal2011} and \citet{deugenioetal2013} in Figure \ref{fig:binfrac} ({\it upper left}).

Because massive galaxies preferentially reside in overdense regions, it is difficult to determine from these small early samples whether or not environment plays a role independent of mass. Two larger IFS surveys have recently published relevant studies. \citet{vealeetal2017b} examined a mass-selected sample of galaxies from MASSIVE \citep{maetal2014}. The majority of these galaxies are central galaxies living in a wide range of halo masses. \citet{vealeetal2017b} do not find a compelling environmental dependence of \lamr\ at fixed $M_*$. \citet{broughetal2017} examine eight clusters observed by SAMI \citep{croometal2012}. By number, their sample is dominated by satellite galaxies, containing only eight brightest-cluster galaxies. They also argue that trends with local overdensity can be explained by the strong dependence on stellar mass. 

We present a bridge between the MASSIVE and SAMI samples. We exploit the large number of IFS cubes afforded by the MaNGA survey \citep[][]{bundyetal2015}, part of the Sloan Digital Sky Survey IV \citep[SDSS IV;][]{blantonetal2017}, to build a sample of galaxies with \lamr\ measurements \citep[][hereafter Paper I]{greeneetal2017}. Roughly $2/3$ of the galaxies in our MaNGA sample are central galaxies, spanning a wide range in inferred host halo mass (and thus environment). We consider the kinematic morphology-density relation of early-type galaxies using a number of different complementary probes of global and local environment, including halo mass, designation as central or satellite, local overdensity, and radial location in the cluster. 

We assume a flat $\Lambda$CDM cosmology with $\Omega_{\rm m}=0.238$, $\Omega_\Lambda=0.762$, H$_0=$100~h$^{-1}$~km~s$^{-1}$~Mpc$^{-1}$, in order to follow the convention of our group catalog \citep{yangetal2007}. Halo masses are defined as $M_{200b}\equiv M(<R_{200b})=200\bar{\rho} \frac{4}{3}\pi R_{200b}^3$ where $R_{200b}$ is the radius at which the mean interior density is equal to 200 times the mean matter density of the universe ($\bar{\rho}$). Stellar mass is denoted $M_{*}$ and has been derived using a Chabrier Initial Mass Function (IMF). We use units in which $h=1$.

\section{Sample and Data}

\subsection{The MaNGA Survey}

MaNGA will ultimately obtain integral-field spectroscopy of 10,000 nearby galaxies with the 2.5m Sloan Foundation Telescope \citep{gunnetal2006} and the BOSS spectrographs \citep{smeeetal2013}. Fibers are joined into 17 hexagonal fiber bundles to perform a multi-object IFS survey \citep{droryetal2015}. Each fiber has a diameter of 2\arcsec, while the bundles range in diameter from 12 to 32\arcsec\ with a 56\% filling factor. The BOSS spectrographs have a wavelength coverage of 3600-10,300\AA\ and a spectral resolution of $\sigma_r \approx 70$~\kms. The relative spectrophotometry is accurate to a few percent \citep{yanetal2016a}. The survey design is described in \citet{yanetal2016b}, the observing strategy in \citet{lawetal2015}, and the data reduction pipeline in \citet{lawetal2016}. 

The MaNGA sample is selected from the NASA-Sloan Atlas (NSA; Blanton M. http://www.nsatlas.org) with redshifts mostly from the SDSS Data Release 7 MAIN galaxy sample \citep{abazajianetal2009}. The sample is built in $i$-band absolute magnitude ($M_i$)-complete shells, with more luminous galaxies observed in more distant $M_i$ shells such that the spatial coverage (in terms of $R_e$) is roughly constant across the sample \citep[][]{yanetal2016b,wakeetal2017}. We focus on the combination of the {\it Primary} sample ($\sim 50\%$ of the total sample) and {\it Secondary} sample ($\sim 40\%$ of the total sample), selected such that $80\%$ of the galaxies in each $M_i$ shell can be covered to $1.5 R_e$ ($2.5 R_e$) by the largest MaNGA IFU for the Primary (Secondary) sample, respectively. To compare different stellar mass bins, it is necessary to reweight the galaxy distributions in the sample to account for the different volumes probed by each mass shell. We apply these volume weights whenever population means are presented.


\subsection{Galaxy Sample}

We work with the MaNGA data derived from the data-release pipeline (DRP) v2.0.1 (MaNGA Product Launch [MPL] 5) sample that are also in the \citet{yangetal2007} group catalog (Y07; updated to DR7).  Y07 use an iterative, adaptive group finder to assign galaxies to halos. Briefly, they first use a friends-of-friends algorithm to identify potential groups, followed by abundance matching to assign likely halo masses to each group based on the total galaxy luminosity. They iterate their group selection based on the initial estimate for halo mass. We select the central galaxy as the most luminous one, while all other galaxies are designated as satellite galaxies. 

We select central and satellite galaxies that reside in halos more massive than $10^{12.5}~h^{-1}~M_{\odot}$ \citep[where the group catalog is complete;][]{yangetal2009} and additionally require that the satellite galaxies have stellar masses $M_*>10^{10}~h^{-2}~M_{\odot}$. We also visually remove all galaxies with spiral structure, focusing only on early-type (E/S0) galaxies (for details see Paper I). The final sample comprises 379 early-type centrals with a minimum stellar mass of $3 \times 10^{10}$~\msun\ and a median stellar mass of $10^{11}$~\msun. There are 159 early-type satellite galaxies with a minimum stellar mass of $10^{10}$~\msun\ and a median stellar mass of $3 \times 10^{10}$~\msun.

\begin{figure*}
\vbox{ 
\vskip -10mm
\hskip +25mm
\includegraphics[width=0.7\textwidth]{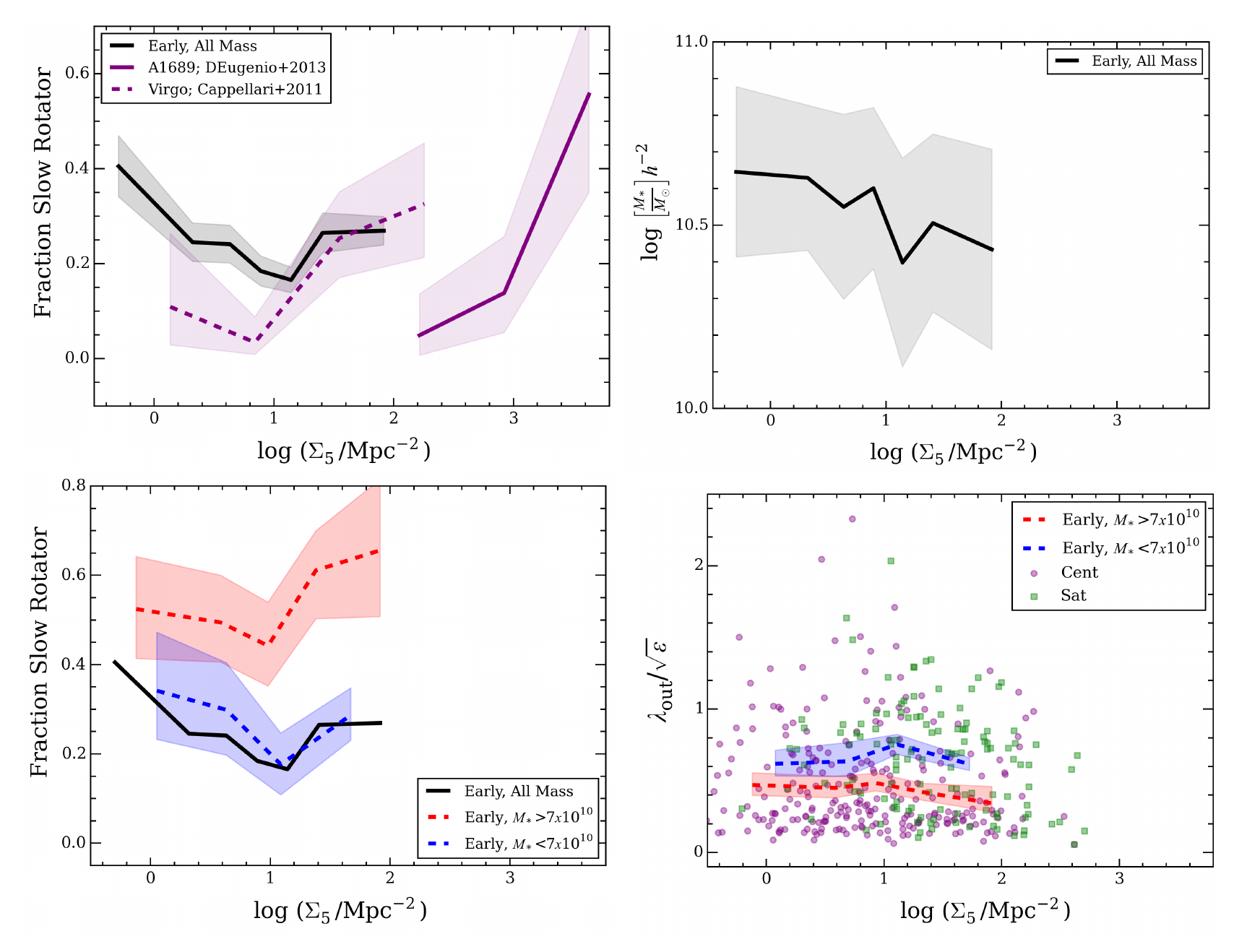}
}
\vskip -0mm
\figcaption[]{
{\it Top Left}: Slow-rotator fraction (with MaNGA weights applied) among the full sample of early-type galaxies as a function of local overdensity $\Sigma_5$. Two clusters from the literature are shown for comparison, Abell 1689 \citep[solid purple;][]{deugenioetal2013} and Virgo \citep[dashed purple;][]{cappellarietal2011}. Our data are binned to contain equal numbers of points, and bins are plotted at the weighted mean value of $M_*$. The shaded regions denote the weighted mean and the error in that mean. There is not a significant trend between slow-rotator fraction and local overdensity.
{\it Top Right}: Stellar mass vs $\Sigma_5$ shows a slight drop in stellar mass towards the highest overdensities due to a preponderance of satellites at high $\Sigma_5$.
{\it Bottom Left}: Mass-weighted slow-rotator fraction as a function of local overdensity. We reweight each galaxy to enforce a uniform mass distribution, and consider a high-mass (red dashed) and low-mass (blue dashed) bin divided at the sample median stellar mass. There is no residual trend with overdensity at fixed mass in these data.
{\it Bottom Right}: The full distribution of \lamout/$\sqrt{\epsilon}$ as a function of $\Sigma_5$. Satellites (squares) and centrals (circles), are both included in the mass-weighted trends. Again, we see no trend as a function of $\Sigma_5$.
\label{fig:binfrac}}
\end{figure*}

\subsection{Identifying Slow Rotators}

The kinematic measurements we use to derive \lamr\ come from the MaNGA Data Analysis Pipeline (DAP), and are measured on Voronoi-binned data \citep{cappellaricopin2003} with a signal-to-noise ratio of at least 10 per 70 km~s$^{-1}$ spectral pixel. The kinematics are measured using the penalized pixel-fitting code pPXF \citep{cappellariemsellem2004}, with emission lines masked. The stellar templates are drawn from the MILES library  \citep{sanchezblazquezetal2006} and are convolved with a Gaussian line-of-sight velocity distribution to derive the velocity and velocity dispersion of the stars. The velocity dispersions are reliable above $\sigma>40$~\kms\ \citep{pennyetal2016}, while the small number of unresolved spaxels are removed from analysis. The central $\sigma$ values range from 50 to 400~\kms, with only 43 galaxies having dispersions $<100$~\kms\ for the sample, so we are not working in this low dispersion regime. An eighth-order additive polynomial is included to account for flux calibration and stellar population mismatch. We adopt the ``DONOTUSE'' flags from the DAP (meaning that there are known catastrophic problems with these data), and flag all bins with $\sigma >500$~\kms\ or $V>400$~\kms. We only keep galaxies for which at least $50\%$ of their spaxels are unflagged.

We calculate \lamr\ within elliptical isophotes. We adopt the position angle and galaxy flattening $\epsilon = 1-b/a$ from the NASA Sloan Atlas \citep[NSA;][]{blantonetal2011}\footnote{http://www.sdss.org/dr13/manga/manga-target-selection/nsa/}. While it is standard in the literature to compare galaxies at \lamre, we show in Paper I that at the spatial resolution of MaNGA these measurements can be biased by $10-50\%$, depending on the input \lamre, with lower values of \lamre\ suffering more severely. As described in detail in Paper I, we adopt the outermost measurement of \lamr\ (\lamout). \lamout\ matches $\lambda$($1.5 R_e$) with $\sim 20\%$ scatter and no bias, but allows us to include galaxies with limited radial coverage. Finally, the inner 2\arcsec\ of data are excluded, since the low spatial resolution of MaNGA tends to lower $\lambda_R$. In Paper I, we use simulations to show that excluding the central region brings the measured $\lambda_R$ value closer to the true value, and we estimate that the residual impact of low spatial resolution leads to at most a systematic increase in slow-rotator fraction of $10\%$ from our measured values.

To determine whether a galaxy is a slow or fast rotator further requires comparison with the intrinsic shape of the galaxy, since the amount of rotation needed to support an oblate galaxy rises with ellipticity. \citet{emsellemetal2007} report an empirically motivated division between slow and fast rotators of \lamre$<0.31 \sqrt{\epsilon}$. For measurements at $\lambda$($1.5 R_e$), we simply scale by the typical ratio of \lamre$/\lambda$($1.5R_e$) from our data, to define slow rotators as those with \lamout$<0.35 \sqrt{\epsilon}$.

\section{The relationship between $\lambda$ and environment}

We now demonstrate that at fixed $M_*$ there is no residual dependence of \lamout\ on either local or global environment. We discuss the different measures of environment used in the analysis, describe how we normalize for $M_*$, and examine \lamout\ as a function of local overdensity at fixed $M_*$.

\subsection{Measures of Environment}

In Paper I we examined \lamout\ in bins of $M_{200b}$, finding no evidence for any residual dependence of \lamout\ on halo mass $M_{200b}$ at fixed $M_*$. Nor was there any significant evidence for differences between central and satellite galaxies at fixed mass. Therefore, in this paper we do not reconsider $M_{200b}$ or the satellite/central distinction. We do note, however, that there is considerable scatter in the halo masses and the central designations in group catalogs \citep[e.g.,][]{campbelletal2015}. It is useful, therefore, to consider local overdensity measures as a complement to the group-based measures of environment.

The first papers investigating links between \lamr\ and environment focused on a local overdensity measurement \citep[e.g.,][]{dressler1980}. The local overdensity measurement employed here ($\Sigma_n$) is simply the number of galaxies (in our case $n=5$) with $M_r>-20.3$~mag, divided by the volume required to enclose that number of neighbors above the magnitude limit. We adopt the $\Sigma_5$ measurements from \citet[][as defined by \citealt{etheringtonthomas2015}]{goddardetal2017}. 

$\Sigma_n$ is a complex measure of environment. For large overdensities, with more than three to five members, the region of measurement falls within the parent halo. However, when the number of members of a group approaches the overdensity measure, the distance to the n$^{\rm th}$ nearest neighbor may fall in a different halo. In this case, the measure becomes more sensitive to very large scale clustering than to local overdensity \citep[][]{muldrewetal2012,wooetal2013}. Furthermore, the physical processes impacting galaxies at group centers (e.g., merging) may be different than those dominating the satellite galaxies living further from the group center (e.g., stripping). We therefore prefer using the radial group position, $R/R_{200b}$, which ensures only intra-group comparisons are made. For comparison with previous literature, we examine both measures. 

\subsection{Controlling for stellar mass}

We employ a simple mass-weighting scheme to ensure flat mass distributions in all environmental bins. In each environment bin, we inversely weight each galaxy based on the number of galaxies at that mass, such that the final mass distribution is flat. We do this in two mass bins, divided at the median stellar mass. We do not employ the MaNGA weights here, since we are looking at renormalized mass distributions.

Throughout the rest of the paper, we will display the weighted slow-rotator fractions as a function of $\Sigma_n$ and $R/R_{200b}$. In these figures, the bin sizes are chosen to ensure a constant number of objects per bin, and are plotted at the weighted bin center. The shaded regions indicate the error in the mean, derived via bootstrapping.

\begin{figure*}
\vbox{ 
\vskip -10mm
\hskip +25mm
\includegraphics[width=0.75\textwidth]{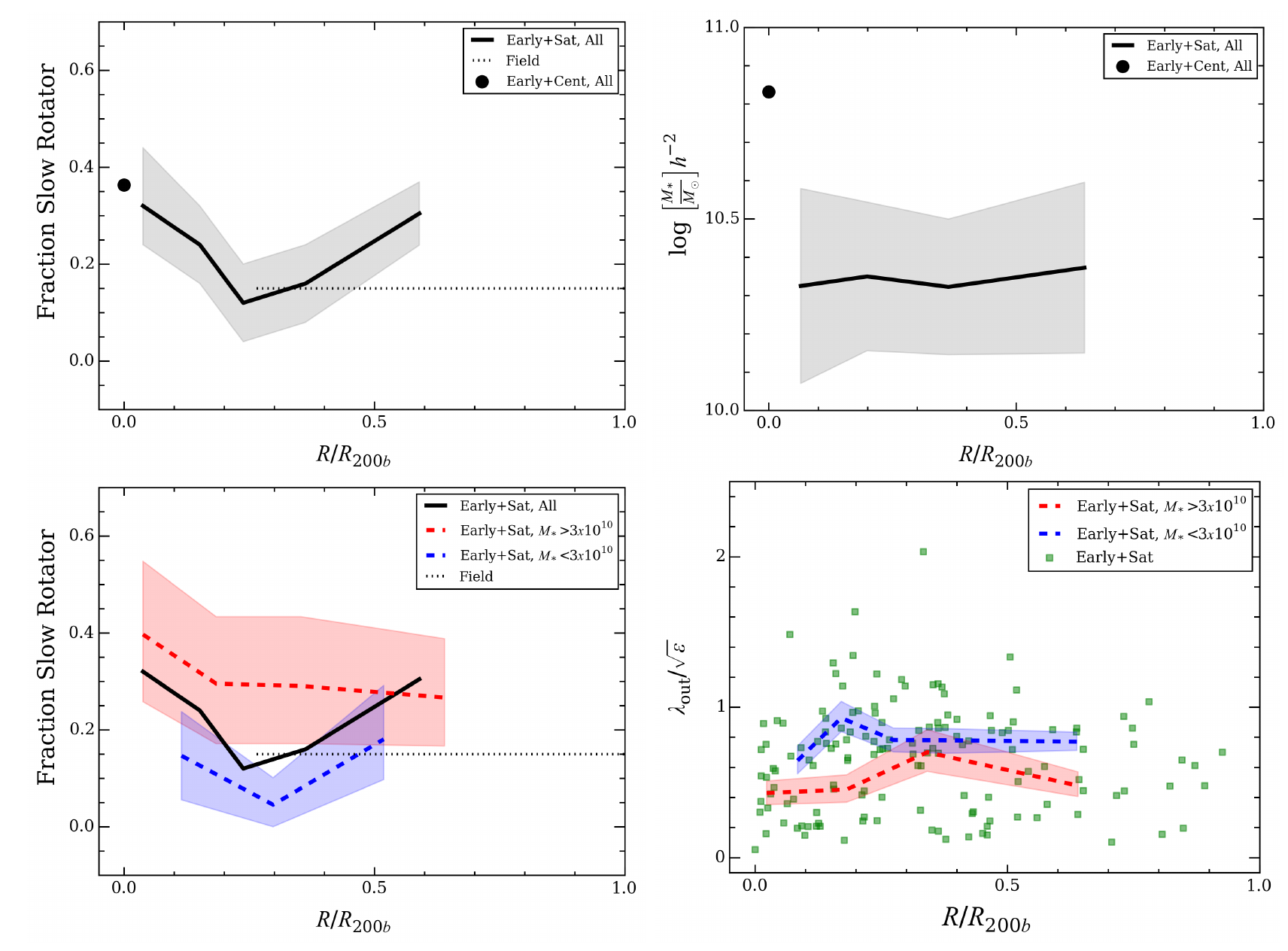}
}
\vskip -0mm
\figcaption[]{
{\it Top Left}: Slow rotator fraction as a function of normalized group radius. Centrals (black point) are at the group center; the rest are satellites (black line).
{\it Top Right}: Weighted-mean stellar mass (and error in the mean) as a function of radius in the group.
{\it Bottom Left}: Mass-weighted slow-rotator fraction as a function of radius. Normalizing for stellar mass, there is no radial dependence of this fraction. We apply the same mass-weighting as in Figure 1 to enforce a uniform mass distribution, and consider a high-mass (red dashed) and low-mass (blue dashed) bin divided at the sample median stellar mass (note it is lower here since we consider only satellite galaxies).
{\it Bottom Right}: Distribution in \lamout/$\sqrt{\epsilon}$ as a function of radial coordinate $R/R_{200b}$ for the satellite galaxies, weighted by mass as in the bottom-right panel. There is a difference in overall \lamout/$\sqrt{\epsilon}$ as a function of stellar mass, but no additional trend with radius.
\label{fig:radfrac}}
\end{figure*}

\subsection{Trends with $\Sigma_5$}

We first examine whether there is a trend between \lamout\ and local overdensity at fixed $M_*$. Figure \ref{fig:binfrac} ({\it top left}) presents two clusters that illustrate the range of results from individual studies of clusters, and the slow-rotator fraction is shown for our full sample of satellite and central galaxies. There is no strong trend with local overdensity in our data, which are roughly consistent with the Virgo results in a similar $\Sigma$ range (although the Virgo results employ $\Sigma_3$ rather than $\Sigma_5$). We see a rising slow rotator fraction at low $\Sigma_5$. This rising fraction reflects the increasing dominance of higher-mass central galaxies at low overdensity in our sample. Figure \ref{fig:binfrac} ({\it top right}) shows that there is a trend (albeit weak) between local overdensity and stellar mass as satellites grow more dominant at higher overdensity. 
To garner reasonable results, we must examine the trends with overdensity at fixed mass. Note the full relationship between stellar mass and local overdensity is complicated for satellites \citep[e.g.,][]{wooetal2013}, and is not fully probed by our early-type, high-mass sample.

In Figure \ref{fig:binfrac} ({\it bottom left}), we impose mass-weighting. While higher-mass galaxies show a higher slow-rotator fraction, no residual trend is seen with $\Sigma_5$ for either bin. Since slow-rotator fraction is assigned as a binary (noisy) division, it is useful to examine the full distribution of \lamout/$\sqrt{\epsilon}$ (Figure \ref{fig:binfrac}; {\it bottom right}). Again, we see a split in samples based on stellar mass, but no residual trend in \lamout/$\sqrt{\epsilon}$ as a function of $\Sigma_5$. We thus turn to examine \lamout\ as a function of $R/R_{200b}$.

\subsection{Trends with $R/R_{200b}$}

$R/R_{200b}$ is complementary to $\Sigma_5$, since the radial distance is always measured within the group halo, while the local overdensity can extend to neighboring halos at low density. Figure \ref{fig:radfrac} ({\it top left}) presents the slow-rotator fraction as a function of $R/R_{200b}$ for all satellites in the sample. The central galaxies by construction are found at the group center, and have a 40\% slow-rotator fraction averaged over all masses. There is no compelling trend in slow-rotator fraction as a function of radius. In terms of mass, the central galaxies are more massive than the satellites, but there is not a strong mass segregation within the satellite galaxies (Figure \ref{fig:radfrac}, {\it top right}). The true radial mass trend is washed out by our narrow range in stellar mass and morphology \citep[e.g.,][]{wooetal2013}.

We then consider the mass-weighted slow-rotator fraction as a function of radius (Figure \ref{fig:radfrac}; {\it bottom left}). Again, the high- and low-mass--weighted samples have higher and lower slow-rotator fractions, respectively, but there is no compelling additional evidence for a radial trend in slow-rotator fraction when we control the distribution in stellar mass. We similarly see a very flat mean \lamout/$\sqrt{\epsilon}$ as a function of radius (Figure \ref{fig:radfrac}; {\it bottom right}). 

In short, there is no evidence that large-scale or local environment plays a driving role in the distribution of \lamout, for an array of environmental measures. Of course, it is possible that a stronger radial trend might be apparent if we could consider only the most massive halos; such analysis required larger samples and should be possible with MaNGA soon.

\section{Discussion \& Summary}

Combining the results from both local environmental indicators, we conclude that angular momentum content at fixed stellar mass is not influenced at a level that we can detect here by the local density or by the radial position within the halo. In Paper I we argued that angular momentum at fixed mass was not influenced by the halo mass or central/satellite distinction. Thus, we conclude that local processes (accretion, star-formation, and merging) determine the angular momentum content of early-type galaxies. This result is in accord with recent results from Illustris \citep{penoyreetal2017} suggesting that only major mergers (not minor ones) have the capacity to significantly alter $\lambda_R$ in the most massive galaxies, while accretion and consumption of gas can alter $\lambda_R$ in lower-mass galaxies. Thus, lower-mass galaxies are spun up by gaseous processes, while massive galaxies are not. In this scenario, stellar mass is the dominant factor setting $\lambda_R$, since even if the total level of accretion and merging is higher in denser environments, stellar mass determines whether $\lambda_R$ is impacted at all by external factors at late times.

Our findings are in accord with two more recent studies \citep{vealeetal2017b,broughetal2017}. We suggest that stellar mass dependence also drives earlier results from individual clusters that found evidence for a rising slow-rotator fraction towards the most overdense regions \citep[e.g.,][]{houghtonetal2013,scottetal2014,deugenioetal2015,cappellari2016}. 

Our work fits into a broader conversation about the role of environment in establishing internal galaxy properties. Galaxy mass functions are a function of environment \citep[e.g.,][]{binggelietal1988}. Once stellar mass is controlled, however, there are only subtle remaining differences as a function of environment for many internal galaxy properties, including morphology and color \citep[e.g.,][]{blantonmoustakas2009,alpaslanetal2015}, star formation rates \citep[][]{wijesingheetal2012}, stellar populations \citep{thomasetal2010} and gradients therein \citep[e.g.,][]{greeneetal2015,goddardetal2017}. As the MaNGA survey progresses, the larger sample size will enable yet more sensitive searches for subtle trends between environment, stellar kinematics, gas content, and stellar populations.

\begin{acknowledgements}

J.E.G. is partially supported by NSF AST-1411642. We thank M. Cappellari for 
useful discussions. We thank the referee for constructive comments that improved this manuscript and we thank Ada Elizabeth Greene Silverman for waiting an extra week to come into this world while her mother submitted the initial manuscript.

Funding for the Sloan Digital Sky Survey IV has been provided by
the Alfred P. Sloan Foundation, the U.S. Department of Energy Office of
Science, and the Participating Institutions. SDSS-IV acknowledges
support and resources from the Center for High-Performance Computing at
the University of Utah. The SDSS web site is www.sdss.org.

SDSS-IV is managed by the Astrophysical Research Consortium for the 
Participating Institutions of the SDSS Collaboration including the 
Brazilian Participation Group, the Carnegie Institution for Science, 
Carnegie Mellon University, the Chilean Participation Group, the French Participation Group, Harvard-Smithsonian Center for Astrophysics, 
Instituto de Astrof\'isica de Canarias, The Johns Hopkins University, 
Kavli Institute for the Physics and Mathematics of the Universe (IPMU) / 
University of Tokyo, Lawrence Berkeley National Laboratory, 
Leibniz Institut f\"ur Astrophysik Potsdam (AIP),  
Max-Planck-Institut f\"ur Astronomie (MPIA Heidelberg), 
Max-Planck-Institut f\"ur Astrophysik (MPA Garching), 
Max-Planck-Institut f\"ur Extraterrestrische Physik (MPE), 
National Astronomical Observatories of China, New Mexico State University, 
New York University, University of Notre Dame, 
Observat\'ario Nacional / MCTI, The Ohio State University, 
Pennsylvania State University, Shanghai Astronomical Observatory, 
United Kingdom Participation Group,
Universidad Nacional Aut\'onoma de M\'exico, University of Arizona, 
University of Colorado Boulder, University of Oxford, University of Portsmouth, 
University of Utah, University of Virginia, University of Washington, University of Wisconsin, 
Vanderbilt University, and Yale University.

This research made use of Marvin, a core Python package and web framework for MaNGA data, developed by Brian Cherinka, Jose Sanchez-Gallego, and Brett Andrews (MaNGA Collaboration, 2017).

\end{acknowledgements}


\begin{thebibliography}{52}
\expandafter\ifx\csname natexlab\endcsname\relax\def\natexlab#1{#1}\fi

\bibitem[{{Abazajian} {et~al.}(2009)}]{abazajianetal2009}
{Abazajian}, K.~N., {et~al.} 2009, \apjs, 182, 543

\bibitem[{{Alpaslan} {et~al.}(2015){Alpaslan}, {Driver}, {Robotham},
  {Obreschkow}, {Andrae}, {Cluver}, {Kelvin}, {Lange}, {Owers}, {Taylor},
  {Andrews}, {Bamford}, {Bland-Hawthorn}, {Brough}, {Brown}, {Colless},
  {Davies}, {Eardley}, {Grootes}, {Hopkins}, {Kennedy}, {Liske},
  {Lara-L{\'o}pez}, {L{\'o}pez-S{\'a}nchez}, {Loveday}, {Madore}, {Mahajan},
  {Meyer}, {Moffett}, {Norberg}, {Penny}, {Pimbblet}, {Popescu}, {Seibert}, \&
  {Tuffs}}]{alpaslanetal2015}
{Alpaslan}, M., {Driver}, S., {Robotham}, A.~S.~G., {Obreschkow}, D., {Andrae},
  E., {Cluver}, M., {Kelvin}, L.~S., {Lange}, R., {Owers}, M., {Taylor}, E.~N.,
  {Andrews}, S.~K., {Bamford}, S., {Bland-Hawthorn}, J., {Brough}, S., {Brown},
  M.~J.~I., {Colless}, M., {Davies}, L.~J.~M., {Eardley}, E., {Grootes}, M.~W.,
  {Hopkins}, A.~M., {Kennedy}, R., {Liske}, J., {Lara-L{\'o}pez}, M.~A.,
  {L{\'o}pez-S{\'a}nchez}, {\'A}.~R., {Loveday}, J., {Madore}, B.~F.,
  {Mahajan}, S., {Meyer}, M., {Moffett}, A., {Norberg}, P., {Penny}, S.,
  {Pimbblet}, K.~A., {Popescu}, C.~C., {Seibert}, M., \& {Tuffs}, R. 2015,
  \mnras, 451, 3249

\bibitem[{{Binggeli} {et~al.}(1988){Binggeli}, {Sandage}, \&
  {Tammann}}]{binggelietal1988}
{Binggeli}, B., {Sandage}, A., \& {Tammann}, G.~A. 1988, \araa, 26, 509

\bibitem[{{Binney}(2005)}]{binney2005}
{Binney}, J. 2005, \mnras, 363, 937

\bibitem[{{Blanton} {et~al.}(2017){Blanton}, {Bershady}, {Abolfathi},
  {Albareti}, {Allende Prieto}, {Almeida}, {Alonso-Garc{\'{\i}}a}, {Anders},
  {Anderson}, {Andrews}, \& et~al.}]{blantonetal2017}
{Blanton}, M.~R., {Bershady}, M.~A., {Abolfathi}, B., {Albareti}, F.~D.,
  {Allende Prieto}, C., {Almeida}, A., {Alonso-Garc{\'{\i}}a}, J., {Anders},
  F., {Anderson}, S.~F., {Andrews}, B., \& et~al. 2017, ArXiv e-prints

\bibitem[{{Blanton} {et~al.}(2011){Blanton}, {Kazin}, {Muna}, {Weaver}, \&
  {Price-Whelan}}]{blantonetal2011}
{Blanton}, M.~R., {Kazin}, E., {Muna}, D., {Weaver}, B.~A., \& {Price-Whelan},
  A. 2011, \aj, 142, 31

\bibitem[{{Blanton} \& {Moustakas}(2009)}]{blantonmoustakas2009}
{Blanton}, M.~R., \& {Moustakas}, J. 2009, \araa, 47, 159

\bibitem[{{Brough} {et~al.}(2017){Brough}, {van de Sande}, {Owers},
  {d'Eugenio}, {Sharp}, {Cortese}, {Scott}, {Croom}, {Bassett}, {Bekki},
  {Bryant}, {Davies}, {Drinkwater}, {Driver}, {Foster}, {Goldstein},
  {Lopez-Sanchez}, {Medling}, {Sweet}, {Taranu}, {Tonini}, {Yi}, {Goodwin},
  {Lawrence}, \& {Richards}}]{broughetal2017}
{Brough}, S., {van de Sande}, J., {Owers}, M.~S., {d'Eugenio}, F., {Sharp},
  R., {Cortese}, L., {Scott}, N., {Croom}, S.~M., {Bassett}, R., {Bekki}, K.,
  {Bryant}, J.~J., {Davies}, R., {Drinkwater}, M.~J., {Driver}, S.~P.,
  {Foster}, C., {Goldstein}, G., {Lopez-Sanchez}, A.~R., {Medling}, A.~M.,
  {Sweet}, S.~M., {Taranu}, D.~S., {Tonini}, C., {Yi}, S.~K., {Goodwin}, M.,
  {Lawrence}, J.~S., \& {Richards}, S.~N. 2017, ArXiv e-prints

\bibitem[{{Bundy} {et~al.}(2015){Bundy}, {Bershady}, {Law}, {Yan}, {Drory},
  {MacDonald}, {Wake}, {Cherinka}, {S{\'a}nchez-Gallego}, {Weijmans}, {Thomas},
  {Tremonti}, {Masters}, {Coccato}, {Diamond-Stanic}, {Arag{\'o}n-Salamanca},
  {Avila-Reese}, {Badenes}, {Falc{\'o}n-Barroso}, {Belfiore}, {Bizyaev},
  {Blanc}, {Bland-Hawthorn}, {Blanton}, {Brownstein}, {Byler}, {Cappellari},
  {Conroy}, {Dutton}, {Emsellem}, {Etherington}, {Frinchaboy}, {Fu}, {Gunn},
  {Harding}, {Johnston}, {Kauffmann}, {Kinemuchi}, {Klaene}, {Knapen},
  {Leauthaud}, {Li}, {Lin}, {Maiolino}, {Malanushenko}, {Malanushenko}, {Mao},
  {Maraston}, {McDermid}, {Merrifield}, {Nichol}, {Oravetz}, {Pan}, {Parejko},
  {Sanchez}, {Schlegel}, {Simmons}, {Steele}, {Steinmetz}, {Thanjavur},
  {Thompson}, {Tinker}, {van den Bosch}, {Westfall}, {Wilkinson}, {Wright},
  {Xiao}, \& {Zhang}}]{bundyetal2015}
{Bundy}, K., {Bershady}, M.~A., {Law}, D.~R., {Yan}, R., {Drory}, N.,
  {MacDonald}, N., {Wake}, D.~A., {Cherinka}, B., {S{\'a}nchez-Gallego}, J.~R.,
  {Weijmans}, A.-M., {Thomas}, D., {Tremonti}, C., {Masters}, K., {Coccato},
  L., {Diamond-Stanic}, A.~M., {Arag{\'o}n-Salamanca}, A., {Avila-Reese}, V.,
  {Badenes}, C., {Falc{\'o}n-Barroso}, J., {Belfiore}, F., {Bizyaev}, D.,
  {Blanc}, G.~A., {Bland-Hawthorn}, J., {Blanton}, M.~R., {Brownstein}, J.~R.,
  {Byler}, N., {Cappellari}, M., {Conroy}, C., {Dutton}, A.~A., {Emsellem}, E.,
  {Etherington}, J., {Frinchaboy}, P.~M., {Fu}, H., {Gunn}, J.~E., {Harding},
  P., {Johnston}, E.~J., {Kauffmann}, G., {Kinemuchi}, K., {Klaene}, M.~A.,
  {Knapen}, J.~H., {Leauthaud}, A., {Li}, C., {Lin}, L., {Maiolino}, R.,
  {Malanushenko}, V., {Malanushenko}, E., {Mao}, S., {Maraston}, C.,
  {McDermid}, R.~M., {Merrifield}, M.~R., {Nichol}, R.~C., {Oravetz}, D.,
  {Pan}, K., {Parejko}, J.~K., {Sanchez}, S.~F., {Schlegel}, D., {Simmons}, A.,
  {Steele}, O., {Steinmetz}, M., {Thanjavur}, K., {Thompson}, B.~A., {Tinker},
  J.~L., {van den Bosch}, R.~C.~E., {Westfall}, K.~B., {Wilkinson}, D.,
  {Wright}, S., {Xiao}, T., \& {Zhang}, K. 2015, \apj, 798, 7

\bibitem[{{Campbell} {et~al.}(2015){Campbell}, {van den Bosch}, {Hearin},
  {Padmanabhan}, {Berlind}, {Mo}, {Tinker}, \& {Yang}}]{campbelletal2015}
{Campbell}, D., {van den Bosch}, F.~C., {Hearin}, A., {Padmanabhan}, N.,
  {Berlind}, A., {Mo}, H.~J., {Tinker}, J., \& {Yang}, X. 2015, \mnras, 452,
  444

\bibitem[{{Cappellari}(2016)}]{cappellari2016}
{Cappellari}, M. 2016, ARAA, in press (arXiv:1602.04267)

\bibitem[{{Cappellari} \& {Copin}(2003)}]{cappellaricopin2003}
{Cappellari}, M., \& {Copin}, Y. 2003, \mnras, 342, 345

\bibitem[{{Cappellari} \& {Emsellem}(2004)}]{cappellariemsellem2004}
{Cappellari}, M., \& {Emsellem}, E. 2004, \pasp, 116, 138

\bibitem[{{Cappellari} {et~al.}(2011){Cappellari}, {Emsellem}, {Krajnovi{\'c}},
  {McDermid}, {Serra}, {Alatalo}, {Blitz}, {Bois}, {Bournaud}, {Bureau},
  {Davies}, {Davis}, {de Zeeuw}, {Khochfar}, {Kuntschner}, {Lablanche},
  {Morganti}, {Naab}, {Oosterloo}, {Sarzi}, {Scott}, {Weijmans}, \&
  {Young}}]{cappellarietal2011}
{Cappellari}, M., {Emsellem}, E., {Krajnovi{\'c}}, D., {McDermid}, R.~M.,
  {Serra}, P., {Alatalo}, K., {Blitz}, L., {Bois}, M., {Bournaud}, F.,
  {Bureau}, M., {Davies}, R.~L., {Davis}, T.~A., {de Zeeuw}, P.~T., {Khochfar},
  S., {Kuntschner}, H., {Lablanche}, P.-Y., {Morganti}, R., {Naab}, T.,
  {Oosterloo}, T., {Sarzi}, M., {Scott}, N., {Weijmans}, A.-M., \& {Young},
  L.~M. 2011, \mnras, 416, 1680

\bibitem[{{Cappellari} {et~al.}(2006)}]{cappellarietal2006}
{Cappellari}, M., {et~al.} 2006, \mnras, 366, 1126

\bibitem[{{Croom} {et~al.}(2012){Croom}, {Lawrence}, {Bland-Hawthorn},
  {Bryant}, {Fogarty}, {Richards}, {Goodwin}, {Farrell}, {Miziarski}, {Heald},
  {Jones}, {Lee}, {Colless}, {Brough}, {Hopkins}, {Bauer}, {Birchall}, {Ellis},
  {Horton}, {Leon-Saval}, {Lewis}, {L{\'o}pez-S{\'a}nchez}, {Min}, {Trinh}, \&
  {Trowland}}]{croometal2012}
{Croom}, S.~M., {Lawrence}, J.~S., {Bland-Hawthorn}, J., {Bryant}, J.~J.,
  {Fogarty}, L., {Richards}, S., {Goodwin}, M., {Farrell}, T., {Miziarski}, S.,
  {Heald}, R., {Jones}, D.~H., {Lee}, S., {Colless}, M., {Brough}, S.,
  {Hopkins}, A.~M., {Bauer}, A.~E., {Birchall}, M.~N., {Ellis}, S., {Horton},
  A., {Leon-Saval}, S., {Lewis}, G., {L{\'o}pez-S{\'a}nchez}, {\'A}.~R., {Min},
  S.-S., {Trinh}, C., \& {Trowland}, H. 2012, \mnras, 421, 872

\bibitem[{{Davies} {et~al.}(1983){Davies}, {Efstathiou}, {Fall}, {Illingworth},
  \& {Schechter}}]{daviesetal1983}
{Davies}, R.~L., {Efstathiou}, G., {Fall}, S.~M., {Illingworth}, G., \&
  {Schechter}, P.~L. 1983, \apj, 266, 41

\bibitem[{{D'Eugenio} {et~al.}(2013){D'Eugenio}, {Houghton}, {Davies}, \&
  {Dalla Bont{\`a}}}]{deugenioetal2013}
{D'Eugenio}, F., {Houghton}, R.~C.~W., {Davies}, R.~L., \& {Dalla Bont{\`a}},
  E. 2013, \mnras, 429, 1258

\bibitem[{{D'Eugenio} {et~al.}(2015){D'Eugenio}, {Houghton}, {Davies}, \&
  {Dalla Bont{\`a}}}]{deugenioetal2015}
---. 2015, \mnras, 451, 827

\bibitem[{{Dressler}(1980)}]{dressler1980}
{Dressler}, A. 1980, \apjs, 42, 565

\bibitem[{{Drory} {et~al.}(2015){Drory}, {MacDonald}, {Bershady}, {Bundy},
  {Gunn}, {Law}, {Smith}, {Stoll}, {Tremonti}, {Wake}, {Yan}, {Weijmans},
  {Byler}, {Cherinka}, {Cope}, {Eigenbrot}, {Harding}, {Holder}, {Huehnerhoff},
  {Jaehnig}, {Jansen}, {Klaene}, {Paat}, {Percival}, \&
  {Sayres}}]{droryetal2015}
{Drory}, N., {MacDonald}, N., {Bershady}, M.~A., {Bundy}, K., {Gunn}, J.,
  {Law}, D.~R., {Smith}, M., {Stoll}, R., {Tremonti}, C.~A., {Wake}, D.~A.,
  {Yan}, R., {Weijmans}, A.~M., {Byler}, N., {Cherinka}, B., {Cope}, F.,
  {Eigenbrot}, A., {Harding}, P., {Holder}, D., {Huehnerhoff}, J., {Jaehnig},
  K., {Jansen}, T.~C., {Klaene}, M., {Paat}, A.~M., {Percival}, J., \&
  {Sayres}, C. 2015, \aj, 149, 77

\bibitem[{{Emsellem} {et~al.}(2011){Emsellem}, {Cappellari}, {Krajnovi{\'c}},
  {Alatalo}, {Blitz}, {Bois}, {Bournaud}, {Bureau}, {Davies}, {Davis}, {de
  Zeeuw}, {Khochfar}, {Kuntschner}, {Lablanche}, {McDermid}, {Morganti},
  {Naab}, {Oosterloo}, {Sarzi}, {Scott}, {Serra}, {van de Ven}, {Weijmans}, \&
  {Young}}]{emsellemetal2011}
{Emsellem}, E., {Cappellari}, M., {Krajnovi{\'c}}, D., {Alatalo}, K., {Blitz},
  L., {Bois}, M., {Bournaud}, F., {Bureau}, M., {Davies}, R.~L., {Davis},
  T.~A., {de Zeeuw}, P.~T., {Khochfar}, S., {Kuntschner}, H., {Lablanche},
  P.-Y., {McDermid}, R.~M., {Morganti}, R., {Naab}, T., {Oosterloo}, T.,
  {Sarzi}, M., {Scott}, N., {Serra}, P., {van de Ven}, G., {Weijmans}, A.-M.,
  \& {Young}, L.~M. 2011, \mnras, 414, 888

\bibitem[{{Emsellem} {et~al.}(2007)}]{emsellemetal2007}
{Emsellem}, E., {et~al.} 2007, \mnras, 379, 401

\bibitem[{{Etherington} \& {Thomas}(2015)}]{etheringtonthomas2015}
{Etherington}, J., \& {Thomas}, D. 2015, \mnras, 451, 660

\bibitem[{{Fogarty} {et~al.}(2014){Fogarty}, {Scott}, {Owers}, {Brough},
  {Croom}, {Pracy}, {Houghton}, {Bland-Hawthorn}, {Colless}, {Davies}, {Jones},
  {Allen}, {Bryant}, {Goodwin}, {Green}, {Konstantopoulos}, {Lawrence},
  {Richards}, {Cortese}, \& {Sharp}}]{fogartyetal2014}
{Fogarty}, L.~M.~R., {Scott}, N., {Owers}, M.~S., {Brough}, S., {Croom}, S.~M.,
  {Pracy}, M.~B., {Houghton}, R.~C.~W., {Bland-Hawthorn}, J., {Colless}, M.,
  {Davies}, R.~L., {Jones}, D.~H., {Allen}, J.~T., {Bryant}, J.~J., {Goodwin},
  M., {Green}, A.~W., {Konstantopoulos}, I.~S., {Lawrence}, J.~S., {Richards},
  S., {Cortese}, L., \& {Sharp}, R. 2014, \mnras, 443, 485

\bibitem[{{Franx} \& {Illingworth}(1990)}]{franxillingworth1990}
{Franx}, M., \& {Illingworth}, G. 1990, \apjl, 359, L41

\bibitem[{{Goddard} {et~al.}(2017){Goddard}, {Thomas}, {Maraston}, {Westfall},
  {Etherington}, {Riffel}, {Mallmann}, {Zheng}, {Argudo-Fern{\'a}ndez},
  {Bershady}, {Bundy}, {Drory}, {Law}, {Yan}, {Wake}, {Weijmans}, {Bizyaev},
  {Brownstein}, {Lane}, {Maiolino}, {Masters}, {Merrifield}, {Nitschelm},
  {Pan}, {Roman-Lopes}, \& {Storchi-Bergmann}}]{goddardetal2017}
{Goddard}, D., {Thomas}, D., {Maraston}, C., {Westfall}, K., {Etherington}, J.,
  {Riffel}, R., {Mallmann}, N.~D., {Zheng}, Z., {Argudo-Fern{\'a}ndez}, M.,
  {Bershady}, M., {Bundy}, K., {Drory}, N., {Law}, D., {Yan}, R., {Wake}, D.,
  {Weijmans}, A., {Bizyaev}, D., {Brownstein}, J., {Lane}, R.~R., {Maiolino},
  R., {Masters}, K., {Merrifield}, M., {Nitschelm}, C., {Pan}, K.,
  {Roman-Lopes}, A., \& {Storchi-Bergmann}, T. 2017, \mnras, 465, 688

\bibitem[{{Greene} {et~al.}(2015){Greene}, {Janish}, {Ma}, {McConnell},
  {Blakeslee}, {Thomas}, \& {Murphy}}]{greeneetal2015}
{Greene}, J.~E., {Janish}, R., {Ma}, C.-P., {McConnell}, N.~J., {Blakeslee},
  J.~P., {Thomas}, J., \& {Murphy}, J.~D. 2015, \apj, 807, 11

\bibitem[{{Greene} {et~al.}(2017){Greene}, {Leauthaud}, \&
  {Emsellem}}]{greeneetal2017}
{Greene}, J.~E., {Leauthaud}, A., \& {Emsellem}, E., e.~a. 2017, \apj,
  submitted, 807, 11

\bibitem[{{Gunn} {et~al.}(2006){Gunn}, {Siegmund}, {Mannery}, {Owen}, {Hull},
  {Leger}, {Carey}, {Knapp}, {York}, {Boroski}, {Kent}, {Lupton}, {Rockosi},
  {Evans}, {Waddell}, {Anderson}, {Annis}, {Barentine}, {Bartoszek}, {Bastian},
  {Bracker}, {Brewington}, {Briegel}, {Brinkmann}, {Brown}, {Carr},
  {Czarapata}, {Drennan}, {Dombeck}, {Federwitz}, {Gillespie}, {Gonzales},
  {Hansen}, {Harvanek}, {Hayes}, {Jordan}, {Kinney}, {Klaene}, {Kleinman},
  {Kron}, {Kresinski}, {Lee}, {Limmongkol}, {Lindenmeyer}, {Long}, {Loomis},
  {McGehee}, {Mantsch}, {Neilsen}, {Neswold}, {Newman}, {Nitta}, {Peoples},
  {Pier}, {Prieto}, {Prosapio}, {Rivetta}, {Schneider}, {Snedden}, \&
  {Wang}}]{gunnetal2006}
{Gunn}, J.~E., {Siegmund}, W.~A., {Mannery}, E.~J., {Owen}, R.~E., {Hull},
  C.~L., {Leger}, R.~F., {Carey}, L.~N., {Knapp}, G.~R., {York}, D.~G.,
  {Boroski}, W.~N., {Kent}, S.~M., {Lupton}, R.~H., {Rockosi}, C.~M., {Evans},
  M.~L., {Waddell}, P., {Anderson}, J.~E., {Annis}, J., {Barentine}, J.~C.,
  {Bartoszek}, L.~M., {Bastian}, S., {Bracker}, S.~B., {Brewington}, H.~J.,
  {Briegel}, C.~I., {Brinkmann}, J., {Brown}, Y.~J., {Carr}, M.~A.,
  {Czarapata}, P.~C., {Drennan}, C.~C., {Dombeck}, T., {Federwitz}, G.~R.,
  {Gillespie}, B.~A., {Gonzales}, C., {Hansen}, S.~U., {Harvanek}, M., {Hayes},
  J., {Jordan}, W., {Kinney}, E., {Klaene}, M., {Kleinman}, S.~J., {Kron},
  R.~G., {Kresinski}, J., {Lee}, G., {Limmongkol}, S., {Lindenmeyer}, C.~W.,
  {Long}, D.~C., {Loomis}, C.~L., {McGehee}, P.~M., {Mantsch}, P.~M.,
  {Neilsen}, Jr., E.~H., {Neswold}, R.~M., {Newman}, P.~R., {Nitta}, A.,
  {Peoples}, Jr., J., {Pier}, J.~R., {Prieto}, P.~S., {Prosapio}, A.,
  {Rivetta}, C., {Schneider}, D.~P., {Snedden}, S., \& {Wang}, S.-i. 2006, \aj,
  131, 2332

\bibitem[{{Houghton} {et~al.}(2013){Houghton}, {Davies}, {D'Eugenio}, {Scott},
  {Thatte}, {Clarke}, {Tecza}, {Salter}, {Fogarty}, \&
  {Goodsall}}]{houghtonetal2013}
{Houghton}, R.~C.~W., {Davies}, R.~L., {D'Eugenio}, F., {Scott}, N., {Thatte},
  N., {Clarke}, F., {Tecza}, M., {Salter}, G.~S., {Fogarty}, L.~M.~R., \&
  {Goodsall}, T. 2013, \mnras, 436, 19

\bibitem[{{Law} {et~al.}(2016){Law}, {Cherinka}, {Yan}, {Andrews}, {Bershady},
  {Bizyaev}, {Blanc}, {Blanton}, {Bolton}, {Brownstein}, {Bundy}, {Chen},
  {Drory}, {D'Souza}, {Fu}, {Jones}, {Kauffmann}, {MacDonald}, {Masters},
  {Newman}, {Parejko}, {S{\'a}nchez-Gallego}, {S{\'a}nchez}, {Schlegel},
  {Thomas}, {Wake}, {Weijmans}, {Westfall}, \& {Zhang}}]{lawetal2016}
{Law}, D.~R., {Cherinka}, B., {Yan}, R., {Andrews}, B.~H., {Bershady}, M.~A.,
  {Bizyaev}, D., {Blanc}, G.~A., {Blanton}, M.~R., {Bolton}, A.~S.,
  {Brownstein}, J.~R., {Bundy}, K., {Chen}, Y., {Drory}, N., {D'Souza}, R.,
  {Fu}, H., {Jones}, A., {Kauffmann}, G., {MacDonald}, N., {Masters}, K.~L.,
  {Newman}, J.~A., {Parejko}, J.~K., {S{\'a}nchez-Gallego}, J.~R.,
  {S{\'a}nchez}, S.~F., {Schlegel}, D.~J., {Thomas}, D., {Wake}, D.~A.,
  {Weijmans}, A.-M., {Westfall}, K.~B., \& {Zhang}, K. 2016, \aj, 152, 83

\bibitem[{{Law} {et~al.}(2015){Law}, {Yan}, {Bershady}, {Bundy}, {Cherinka},
  {Drory}, {MacDonald}, {S{\'a}nchez-Gallego}, {Wake}, {Weijmans}, {Blanton},
  {Klaene}, {Moran}, {Sanchez}, \& {Zhang}}]{lawetal2015}
{Law}, D.~R., {Yan}, R., {Bershady}, M.~A., {Bundy}, K., {Cherinka}, B.,
  {Drory}, N., {MacDonald}, N., {S{\'a}nchez-Gallego}, J.~R., {Wake}, D.~A.,
  {Weijmans}, A.-M., {Blanton}, M.~R., {Klaene}, M.~A., {Moran}, S.~M.,
  {Sanchez}, S.~F., \& {Zhang}, K. 2015, \aj, 150, 19

\bibitem[{{Ma} {et~al.}(2014){Ma}, {Greene}, {McConnell}, {Janish},
  {Blakeslee}, {Thomas}, \& {Murphy}}]{maetal2014}
{Ma}, C.-P., {Greene}, J.~E., {McConnell}, N., {Janish}, R., {Blakeslee},
  J.~P., {Thomas}, J., \& {Murphy}, J.~D. 2014, \apj, 795, 158

\bibitem[{{Muldrew} {et~al.}(2012){Muldrew}, {Croton}, {Skibba}, {Pearce},
  {Ann}, {Baldry}, {Brough}, {Choi}, {Conselice}, {Cowan}, {Gallazzi}, {Gray},
  {Gr{\"u}tzbauch}, {Li}, {Park}, {Pilipenko}, {Podgorzec}, {Robotham},
  {Wilman}, {Yang}, {Zhang}, \& {Zibetti}}]{muldrewetal2012}
{Muldrew}, S.~I., {Croton}, D.~J., {Skibba}, R.~A., {Pearce}, F.~R., {Ann},
  H.~B., {Baldry}, I.~K., {Brough}, S., {Choi}, Y.-Y., {Conselice}, C.~J.,
  {Cowan}, N.~B., {Gallazzi}, A., {Gray}, M.~E., {Gr{\"u}tzbauch}, R., {Li},
  I.-H., {Park}, C., {Pilipenko}, S.~V., {Podgorzec}, B.~J., {Robotham},
  A.~S.~G., {Wilman}, D.~J., {Yang}, X., {Zhang}, Y., \& {Zibetti}, S. 2012,
  \mnras, 419, 2670

\bibitem[{{Naab} {et~al.}(2014){Naab}, {Oser}, {Emsellem}, {Cappellari},
  {Krajnovi{\'c}}, {McDermid}, {Alatalo}, {Bayet}, {Blitz}, {Bois}, {Bournaud},
  {Bureau}, {Crocker}, {Davies}, {Davis}, {de Zeeuw}, {Duc}, {Hirschmann},
  {Johansson}, {Khochfar}, {Kuntschner}, {Morganti}, {Oosterloo}, {Sarzi},
  {Scott}, {Serra}, {Ven}, {Weijmans}, \& {Young}}]{naabetal2014}
{Naab}, T., {Oser}, L., {Emsellem}, E., {Cappellari}, M., {Krajnovi{\'c}}, D.,
  {McDermid}, R.~M., {Alatalo}, K., {Bayet}, E., {Blitz}, L., {Bois}, M.,
  {Bournaud}, F., {Bureau}, M., {Crocker}, A., {Davies}, R.~L., {Davis}, T.~A.,
  {de Zeeuw}, P.~T., {Duc}, P.-A., {Hirschmann}, M., {Johansson}, P.~H.,
  {Khochfar}, S., {Kuntschner}, H., {Morganti}, R., {Oosterloo}, T., {Sarzi},
  M., {Scott}, N., {Serra}, P., {Ven}, G.~v.~d., {Weijmans}, A., \& {Young},
  L.~M. 2014, \mnras, 444, 3357

\bibitem[{{Oliva-Altamirano} {et~al.}(2017){Oliva-Altamirano}, {Brough},
  {Tran}, {Jimmy}, {Miller}, {Bremer}, {Phillipps}, {Sharp}, {Colless},
  {Lara-L{\'o}pez}, {L{\'o}pez-S{\'a}nchez}, {Pimbblet}, {Kafle}, \&
  {Couch}}]{olivaetal2017}
{Oliva-Altamirano}, P., {Brough}, S., {Tran}, K.-V., {Jimmy}, {Miller}, C.,
  {Bremer}, M.~N., {Phillipps}, S., {Sharp}, R., {Colless}, M.,
  {Lara-L{\'o}pez}, M.~A., {L{\'o}pez-S{\'a}nchez}, {\'A}.~R., {Pimbblet}, K.,
  {Kafle}, P.~R., \& {Couch}, W.~J. 2017, \aj, 153, 89

\bibitem[{{Penny} {et~al.}(2016){Penny}, {Masters}, {Weijmans}, {Westfall},
  {Bershady}, {Bundy}, {Drory}, {Falc{\'o}n-Barroso}, {Law}, {Nichol},
  {Thomas}, {Bizyaev}, {Brownstein}, {Freischlad}, {Gaulme}, {Grabowski},
  {Kinemuchi}, {Malanushenko}, {Malanushenko}, {Oravetz}, {Roman-Lopes}, {Pan},
  {Simmons}, \& {Wake}}]{pennyetal2016}
{Penny}, S.~J., {Masters}, K.~L., {Weijmans}, A.-M., {Westfall}, K.~B.,
  {Bershady}, M.~A., {Bundy}, K., {Drory}, N., {Falc{\'o}n-Barroso}, J., {Law},
  D., {Nichol}, R.~C., {Thomas}, D., {Bizyaev}, D., {Brownstein}, J.~R.,
  {Freischlad}, G., {Gaulme}, P., {Grabowski}, K., {Kinemuchi}, K.,
  {Malanushenko}, E., {Malanushenko}, V., {Oravetz}, D., {Roman-Lopes}, A.,
  {Pan}, K., {Simmons}, A., \& {Wake}, D.~A. 2016, \mnras, 462, 3955

\bibitem[{{Penoyre} {et~al.}(2017){Penoyre}, {Moster}, {Sijacki}, \&
  {Genel}}]{penoyreetal2017}
{Penoyre}, Z., {Moster}, B.~P., {Sijacki}, D., \& {Genel}, S. 2017, MNRAS,
  submitted (arXiv:1703.00545)

\bibitem[{{S{\'a}nchez-Bl{\'a}zquez} {et~al.}(2006){S{\'a}nchez-Bl{\'a}zquez},
  {Gorgas}, {Cardiel}, \& {Gonz{\'a}lez}}]{sanchezblazquezetal2006}
{S{\'a}nchez-Bl{\'a}zquez}, P., {Gorgas}, J., {Cardiel}, N., \& {Gonz{\'a}lez},
  J.~J. 2006, \aap, 457, 787

\bibitem[{{Scott} {et~al.}(2014){Scott}, {Davies}, {Houghton}, {Cappellari},
  {Graham}, \& {Pimbblet}}]{scottetal2014}
{Scott}, N., {Davies}, R.~L., {Houghton}, R.~C.~W., {Cappellari}, M., {Graham},
  A.~W., \& {Pimbblet}, K.~A. 2014, \mnras, 441, 274

\bibitem[{{Smee} {et~al.}(2013){Smee}, {Gunn}, {Uomoto}, {Roe}, {Schlegel},
  {Rockosi}, {Carr}, {Leger}, {Dawson}, {Olmstead}, {Brinkmann}, {Owen},
  {Barkhouser}, {Honscheid}, {Harding}, {Long}, {Lupton}, {Loomis}, {Anderson},
  {Annis}, {Bernardi}, {Bhardwaj}, {Bizyaev}, {Bolton}, {Brewington}, {Briggs},
  {Burles}, {Burns}, {Castander}, {Connolly}, {Davenport}, {Ebelke}, {Epps},
  {Feldman}, {Friedman}, {Frieman}, {Heckman}, {Hull}, {Knapp}, {Lawrence},
  {Loveday}, {Mannery}, {Malanushenko}, {Malanushenko}, {Merrelli}, {Muna},
  {Newman}, {Nichol}, {Oravetz}, {Pan}, {Pope}, {Ricketts}, {Shelden},
  {Sandford}, {Siegmund}, {Simmons}, {Smith}, {Snedden}, {Schneider},
  {SubbaRao}, {Tremonti}, {Waddell}, \& {York}}]{smeeetal2013}
{Smee}, S.~A., {Gunn}, J.~E., {Uomoto}, A., {Roe}, N., {Schlegel}, D.,
  {Rockosi}, C.~M., {Carr}, M.~A., {Leger}, F., {Dawson}, K.~S., {Olmstead},
  M.~D., {Brinkmann}, J., {Owen}, R., {Barkhouser}, R.~H., {Honscheid}, K.,
  {Harding}, P., {Long}, D., {Lupton}, R.~H., {Loomis}, C., {Anderson}, L.,
  {Annis}, J., {Bernardi}, M., {Bhardwaj}, V., {Bizyaev}, D., {Bolton}, A.~S.,
  {Brewington}, H., {Briggs}, J.~W., {Burles}, S., {Burns}, J.~G., {Castander},
  F.~J., {Connolly}, A., {Davenport}, J.~R.~A., {Ebelke}, G., {Epps}, H.,
  {Feldman}, P.~D., {Friedman}, S.~D., {Frieman}, J., {Heckman}, T., {Hull},
  C.~L., {Knapp}, G.~R., {Lawrence}, D.~M., {Loveday}, J., {Mannery}, E.~J.,
  {Malanushenko}, E., {Malanushenko}, V., {Merrelli}, A.~J., {Muna}, D.,
  {Newman}, P.~R., {Nichol}, R.~C., {Oravetz}, D., {Pan}, K., {Pope}, A.~C.,
  {Ricketts}, P.~G., {Shelden}, A., {Sandford}, D., {Siegmund}, W., {Simmons},
  A., {Smith}, D.~S., {Snedden}, S., {Schneider}, D.~P., {SubbaRao}, M.,
  {Tremonti}, C., {Waddell}, P., \& {York}, D.~G. 2013, \aj, 146, 32

\bibitem[{{Thomas} {et~al.}(2010){Thomas}, {Maraston}, {Schawinski}, {Sarzi},
  \& {Silk}}]{thomasetal2010}
{Thomas}, D., {Maraston}, C., {Schawinski}, K., {Sarzi}, M., \& {Silk}, J.
  2010, \mnras, 404, 1775

\bibitem[{{Veale} {et~al.}(2017{\natexlab{b}}){Veale}, {Ma}, {Greene},
  {Thomas}, {Blakeslee}, {McConnell}, {Walsh}, \& {Ito}}]{vealeetal2017b}
{Veale}, M., {Ma}, C.-P., {Greene}, J.~E., {Thomas}, J., {Blakeslee}, J.~P.,
  {McConnell}, N., {Walsh}, J.~L., \& {Ito}, J. 2017{\natexlab{b}}, \mnras,
  471, 1428

\bibitem[{{Veale} {et~al.}(2017{\natexlab{a}}){Veale}, {Ma}, {Thomas},
  {Greene}, {McConnell}, {Walsh}, {Ito}, {Blakeslee}, \&
  {Janish}}]{vealeetal2017a}
{Veale}, M., {Ma}, C.-P., {Thomas}, J., {Greene}, J.~E., {McConnell}, N.~J.,
  {Walsh}, J., {Ito}, J., {Blakeslee}, J.~P., \& {Janish}, R.
  2017{\natexlab{a}}, \mnras, 464, 356

\bibitem[{{Wake} {et~al.}(2017){Wake}, {Bundy}, {Diamond-Stanic}, {Yan},
  {Blanton}, {Bershady}, {S{\'a}nchez-Gallego}, {Drory}, {Jones}, {Kauffmann},
  {Law}, {Li}, {MacDonald}, {Masters}, {Thomas}, {Tinker}, {Weijmans}, \&
  {Brownstein}}]{wakeetal2017}
{Wake}, D.~A., {Bundy}, K., {Diamond-Stanic}, A.~M., {Yan}, R., {Blanton},
  M.~R., {Bershady}, M.~A., {S{\'a}nchez-Gallego}, J.~R., {Drory}, N., {Jones},
  A., {Kauffmann}, G., {Law}, D.~R., {Li}, C., {MacDonald}, N., {Masters}, K.,
  {Thomas}, D., {Tinker}, J., {Weijmans}, A.-M., \& {Brownstein}, J.~R. 2017,
  ArXiv e-prints

\bibitem[{{Wijesinghe} {et~al.}(2012){Wijesinghe}, {Hopkins}, {Brough},
  {Taylor}, {Norberg}, {Bauer}, {Brown}, {Cameron}, {Conselice}, {Croom},
  {Driver}, {Grootes}, {Jones}, {Kelvin}, {Loveday}, {Pimbblet}, {Popescu},
  {Prescott}, {Sharp}, {Baldry}, {Sadler}, {Liske}, {Robotham}, {Bamford},
  {Bland-Hawthorn}, {Gunawardhana}, {Meyer}, {Parkinson}, {Drinkwater},
  {Peacock}, \& {Tuffs}}]{wijesingheetal2012}
{Wijesinghe}, D.~B., {Hopkins}, A.~M., {Brough}, S., {Taylor}, E.~N.,
  {Norberg}, P., {Bauer}, A., {Brown}, M.~J.~I., {Cameron}, E., {Conselice},
  C.~J., {Croom}, S., {Driver}, S., {Grootes}, M.~W., {Jones}, D.~H., {Kelvin},
  L., {Loveday}, J., {Pimbblet}, K.~A., {Popescu}, C.~C., {Prescott}, M.,
  {Sharp}, R., {Baldry}, I., {Sadler}, E.~M., {Liske}, J., {Robotham},
  A.~S.~G., {Bamford}, S., {Bland-Hawthorn}, J., {Gunawardhana}, M., {Meyer},
  M., {Parkinson}, H., {Drinkwater}, M.~J., {Peacock}, J., \& {Tuffs}, R. 2012,
  \mnras, 423, 3679

\bibitem[{{Woo} {et~al.}(2013){Woo}, {Dekel}, {Faber}, {Noeske}, {Koo},
  {Gerke}, {Cooper}, {Salim}, {Dutton}, {Newman}, {Weiner}, {Bundy}, {Willmer},
  {Davis}, \& {Yan}}]{wooetal2013}
{Woo}, J., {Dekel}, A., {Faber}, S.~M., {Noeske}, K., {Koo}, D.~C., {Gerke},
  B.~F., {Cooper}, M.~C., {Salim}, S., {Dutton}, A.~A., {Newman}, J., {Weiner},
  B.~J., {Bundy}, K., {Willmer}, C.~N.~A., {Davis}, M., \& {Yan}, R. 2013,
  \mnras, 428, 3306

\bibitem[{{Yan} {et~al.}(2016{\natexlab{a}}){Yan}, {Bundy}, {Law}, {Bershady},
  {Andrews}, {Cherinka}, {Diamond-Stanic}, {Drory}, {MacDonald},
  {S{\'a}nchez-Gallego}, {Thomas}, {Wake}, {Weijmans}, {Westfall}, {Zhang},
  {Arag{\'o}n-Salamanca}, {Belfiore}, {Bizyaev}, {Blanc}, {Blanton},
  {Brownstein}, {Cappellari}, {D'Souza}, {Emsellem}, {Fu}, {Gaulme}, {Graham},
  {Goddard}, {Gunn}, {Harding}, {Jones}, {Kinemuchi}, {Li}, {Li}, {Maiolino},
  {Mao}, {Maraston}, {Masters}, {Merrifield}, {Oravetz}, {Pan}, {Parejko},
  {Sanchez}, {Schlegel}, {Simmons}, {Thanjavur}, {Tinker}, {Tremonti}, {van den
  Bosch}, \& {Zheng}}]{yanetal2016b}
{Yan}, R., {Bundy}, K., {Law}, D.~R., {Bershady}, M.~A., {Andrews}, B.,
  {Cherinka}, B., {Diamond-Stanic}, A.~M., {Drory}, N., {MacDonald}, N.,
  {S{\'a}nchez-Gallego}, J.~R., {Thomas}, D., {Wake}, D.~A., {Weijmans}, A.-M.,
  {Westfall}, K.~B., {Zhang}, K., {Arag{\'o}n-Salamanca}, A., {Belfiore}, F.,
  {Bizyaev}, D., {Blanc}, G.~A., {Blanton}, M.~R., {Brownstein}, J.,
  {Cappellari}, M., {D'Souza}, R., {Emsellem}, E., {Fu}, H., {Gaulme}, P.,
  {Graham}, M.~T., {Goddard}, D., {Gunn}, J.~E., {Harding}, P., {Jones}, A.,
  {Kinemuchi}, K., {Li}, C., {Li}, H., {Maiolino}, R., {Mao}, S., {Maraston},
  C., {Masters}, K., {Merrifield}, M.~R., {Oravetz}, D., {Pan}, K., {Parejko},
  J.~K., {Sanchez}, S.~F., {Schlegel}, D., {Simmons}, A., {Thanjavur}, K.,
  {Tinker}, J., {Tremonti}, C., {van den Bosch}, R., \& {Zheng}, Z.
  2016{\natexlab{a}}, \aj, 152, 197

\bibitem[{{Yan} {et~al.}(2016{\natexlab{b}}){Yan}, {Tremonti}, {Bershady},
  {Law}, {Schlegel}, {Bundy}, {Drory}, {MacDonald}, {Bizyaev}, {Blanc},
  {Blanton}, {Cherinka}, {Eigenbrot}, {Gunn}, {Harding}, {Hogg},
  {S{\'a}nchez-Gallego}, {S{\'a}nchez}, {Wake}, {Weijmans}, {Xiao}, \&
  {Zhang}}]{yanetal2016a}
{Yan}, R., {Tremonti}, C., {Bershady}, M.~A., {Law}, D.~R., {Schlegel}, D.~J.,
  {Bundy}, K., {Drory}, N., {MacDonald}, N., {Bizyaev}, D., {Blanc}, G.~A.,
  {Blanton}, M.~R., {Cherinka}, B., {Eigenbrot}, A., {Gunn}, J.~E., {Harding},
  P., {Hogg}, D.~W., {S{\'a}nchez-Gallego}, J.~R., {S{\'a}nchez}, S.~F.,
  {Wake}, D.~A., {Weijmans}, A.-M., {Xiao}, T., \& {Zhang}, K.
  2016{\natexlab{b}}, \aj, 151, 8

\bibitem[{{Yang} {et~al.}(2009){Yang}, {Mo}, \& {van den Bosch}}]{yangetal2009}
{Yang}, X., {Mo}, H.~J., \& {van den Bosch}, F.~C. 2009, \apj, 695, 900

\bibitem[{{Yang} {et~al.}(2007){Yang}, {Mo}, {van den Bosch}, {Pasquali}, {Li},
  \& {Barden}}]{yangetal2007}
{Yang}, X., {Mo}, H.~J., {van den Bosch}, F.~C., {Pasquali}, A., {Li}, C., \&
  {Barden}, M. 2007, \apj, 671, 153

\end{thebibliography}

\end{document}